\documentclass[lettersize,journal]{IEEEtran}
\usepackage{amsmath,amsfonts,mathtools}
\usepackage{algorithmic}
\usepackage{algorithm}
\usepackage{array}
\usepackage[caption=false,font=normalsize,labelfont=sf,textfont=sf]{subfig}
\usepackage{textcomp}
\usepackage{stfloats}
\usepackage{url}
\usepackage{verbatim}
\usepackage{graphicx}
\usepackage{cite}
\usepackage{color}
\usepackage{bm}

\usepackage{tabularx,booktabs}
\usepackage[table,xcdraw]{xcolor}
\DeclarePairedDelimiter\abs{\lvert}{\rvert}

\begin{document}

\title{Self-supervised Learning of Audio Representations from Audio-Visual Data using Spatial Alignment}

\author{Shanshan Wang,~\IEEEmembership{Student Member,~IEEE}, Archontis Politis,~\IEEEmembership{Member,~IEEE}, Annamaria Mesaros,~\IEEEmembership{Member,~IEEE}, Tuomas Virtanen ~\IEEEmembership{Fellow,~IEEE}
       
\thanks{Manuscript received Jan 15, 2022; revised , 2022.\\
The authors wish to thank CSC-IT Centre of Science Ltd., Finland,  for providing computational resources. 
A. M. received funding from Academy of Finland grant 332063 ``Teaching machines to listen"}
\thanks{S. Wang, A. Politis, A. Mesaros, and T. Virtanen are with the Faculty of Information Technology and Communication Sciences, Tampere University, Finland, e-mail: \{shanshan.wang, archontis.politis, annamaria.mesaros, tuomas.virtanen\}@tuni.fi}
}

\markboth{Journal of Selected Topics in Signal Processing ,~Vol.~, No.~,Month~2022}%
{Wang \MakeLowercase{\textit{et al.}}: Self-supervised Learning of Audio Representations from Audio-Visual Data using Spatial Alignment}

\maketitle

\begin{abstract}
Learning from audio-visual data offers many possibilities to express correspondence between the audio and visual content, similar to the human perception that relates aural and visual information. 
In this work, we present a method for self-supervised representation learning  based on audio-visual spatial alignment (AVSA), a more sophisticated alignment task than the audio-visual correspondence (AVC). In addition to the correspondence, AVSA also learns from the spatial location of acoustic and visual content. Based on 360\textdegree \ video and Ambisonics audio, we propose selection of visual objects using object detection, and beamforming of the audio signal towards the detected objects, attempting to learn the spatial alignment between objects and the sound they produce. We investigate the use of spatial audio features to represent the audio input, and different audio formats: Ambisonics, mono, and stereo. Experimental results show a 10\% improvement on AVSA for the first order ambisonics intensity vector (FOA-IV) in comparison with log-mel spectrogram features; the addition of object-oriented crops also brings significant performance increases for the human action recognition downstream task. A number of audio-only downstream tasks are devised for testing the effectiveness of the learnt audio feature representation, obtaining performance comparable to state-of-the-art methods on acoustic scene classification from ambisonic and binaural audio.

\end{abstract}


\section{Introduction}

\IEEEPARstart{C}{urrent} state-of-the-art methods in audio classification follow the trend from the visual domain in using large models that require large datasets for training. Unlike in the visual domain, large annotated audio datasets are scarce, due to the high annotation cost (time and effort-wise). On the other hand, audio data is relatively easy and fast to record. It is therefore possible to produce large datasets without annotations.

Self-supervised learning has become a common approach  to learn representations using large datasets containing unlabeled data. 
Self-supervision refers to methods that learn representations of the data using so-called proxy learning tasks, in which the learning process is guided by patterns in the data and the main goal is to learn mappings from input examples to low-dimensional representations. 
These representations, called \emph{embeddings}, can be later used as features in downstream tasks, or the trained network can be fine-tuned using the downstream task data and setup. 

Methods for learning embeddings for audio can be based on audio only or different combinations of audio and other modalities, e.g., audio-visual data, audio with corresponding textual descriptions, or tags. 
For example, language-agnostic speech embeddings learned using only speech signals were successfully used in emotion classification on languages different from that used in pretraining \cite{nandan2020language}. 
In a similar manner, Fonseca et al. \cite{fonseca2021unsupervised} 
trained a deep neural to learn the association between different views of the same time-frequency representation of a signal, in other words to be insensitive to transformations applied by different augmentation methods to the original data. Trained on FSD18k noisy dataset \cite{fonseca2019learning} and tested on the in-domain task of sound event classification, the pretrained model exceeded or at least attained the same performance as a baseline system trained in a supervised  manner \cite{fonseca2021unsupervised}, showing that robust learning is possible without use of explicit labels. Another recent self-supervised learning approach for audio based on unsupervised sound source separation learns the association between the mixture audio and one separated track of the same audio segment \cite{fonseca2021self-sup}. Trained on AudioSet and tested on the in-domain task of audio tagging, the method was shown to learn useful representations even with imperfectly separated tracks. 
These methods prove that semantic structure in the data can be learned without explicit supervision using only the audio signal itself.

The scientific literature also contains a number of approaches using audio-visual data that learn to distinguish visual and aural signals with an appropriately designed learning objective that promotes associating the two. A popular strategy is the use of audio-visual correspondence as proxy task, which has proved to be a powerful criterion for unsupervised or self-supervised representation learning \cite{DBLP:journals/corr/ArandjelovicZ17, harwath2018jointly, owens2018audio, 8682475}, and useful in various downstream audio-visual or audio-only classification or recognition tasks \cite{DBLP:journals/corr/ArandjelovicZ17, Wang2021}. While these works can leverage spatial visual information on the video, as shown by visualizing sound activity on the visual objects that produce it \cite{senocak2018learning}, they work with monophonic audio, dismissing spatial information in audio, if more than one audio channels are available. 

Only a handful of audio-visual studies have exploited the spatial information in multichannel recordings that may reveal directions of sound events, line-of-sight sources, reverberant conditions, and other acoustic characteristics of the recorded scene. The work of Gan et al. \cite{gan2019self} transfers knowledge from videos of moving vehicles to train a model to localize them using only the audio from a stereo microphone, while Valverde et al. \cite{valverde2021there} attempt a similar task including additional modalities such as depth maps and thermal maps, and a larger microphone array.  Vasudevan et al. \cite{vasudevan2020semantic} use object detection and depth maps from 360\textdegree \ video as supervision for an audio network using four pairs of binaural microphones. They study a multi-task setting that includes sound event detection and localization (SELD), generating 360\textdegree \ depth maps from audio, and generating rotated binaural signals at unseen orientations. Irie et al. \cite{irie2019seeing} combine video along with corresponding spatial acoustic activity visualizations, obtained with microphone arrays, in order to infer object segmentation masks using only audio at run-time. Perez et al. \cite{perez2020audio} follow a similar strategy for sound event detection showing benefits over using only monophonic audio. They additionally use the audiovisual input to transfer knowledge successfully into monophonic audio classification.

Two recent works have been especially focused on leveraging information from spatial audio cues together with visual data in order to obtain powerful embeddings that can be useful in various downstream visual or audio tasks. In \cite{yang2020telling} ASMR videos from YouTube with corresponding 2-channel binaural audio are fed to a neural network with their left and right channels either in the correct order or flipped. The network, trying to learn to differentiate the correct from incorrect samples, learns spatial representations that prove useful in further audiovisual tasks, surpassing non-spatial audio versions. Morgado et al. \cite{NEURIPS2020_328e5d4c} extend this idea to 360\textdegree \ videos with 4-channel ambisonic audio. 
They train a network to distinguish whether crops in a 360\textdegree \  video frame are spatially aligned with the corresponding ambisonic audio, to learn a representation between spatially aligned views from 360\textdegree \ video and 360\textdegree \ spatial audio, through a contrastive learning approach.

The audio-visual alignment performed in \cite{NEURIPS2020_328e5d4c} was shown to learn useful embeddings for a number of downstream video classification task. However, its focus on the visual side means that the learned audio embeddings were only used when testing the in-domain audio-visual alignment tasks, therefore no explicit audio downstream task was investigated. Moreover, the ambisonic format used in the study is not common in audio datasets, unless they specifically target spatial audio rendering, sound source tracking, or localization. 
The feature representation for the audio input was the log-mel spectrum, even though the Ambisonics format provides the opportunity for spatial audio features, which may be more suitable for the intended alignment task. 
The learning method in \cite{NEURIPS2020_328e5d4c} uses random crops from the video frames. For the purpose of audio-visual alignment, random crops may not contain relevant information with respect to the audio, and therefore a more careful selection of the content may be beneficial. 

In this work, we propose a system that learns from 360\textdegree \  audio and video data through spatial alignment. We build on the method proposed by Morgado et al. \cite{NEURIPS2020_328e5d4c} by proposing a number of specific audio processing steps. 
We focus on the audio-visual spatial alignment because we presume that the audio-visual correspondence with addition of spatial audio-visual alignment  should provide more powerful embeddings than only correspondence, including for audio-only tasks. Even though tasks, applications, and downstream datasets may not be based on 360\textdegree \ video and audio, this format offers a powerful learning strategy based on a full field of view (everything that makes a sound is visible if not occluded or too far). Furthermore, the Ambisonics format can be transformed straightforwardly into "lower-dimensionality" practical formats, such as stereo or mono, but using its full spatial diversity during training will transfer knowledge to the downstream tasks even on these spatially limited formats. We therefore consider worthy of investigation how the AVSA training strategy applies to audio tasks, and how effective audio embeddings can be learned for the different audio formats.

The contribution of this work are as follows: (1) We combine audio beamforming with visual object detection in order to create a strong spatial correspondence between the audio and video modalities, and therefore feed the learning process with a correspondence between visual objects in the scene and the sounds they produce; (2) We represent the audio signal using spatial audio features, to provide an explicit representation of its spatial content; (3) We investigate the use of stereo and mono (beamformed) audio formats in combination with object detection and spatial features, and examine the effect of this transformation on both in-domain and out-of-domain
downstream tasks. We show the usefulness of our approach in a number of audio and video downstream tasks, including the in-domain AVC and AVSA, human action recognition, acoustic scene classification with ambisonic audio using the Eigenscape dataset \cite{app7111204}, and with binaural audio using the TAU Audio-Visual Urban Scenes 2021 dataset, and audio-visual scene classification \cite{Wang_ICASSP2021}.

The rest of this paper is organized as follows: Section \ref{sec:related} introduces in more detail the methods for learning from audio-visual data that serve as background to our work. Section \ref{sec:spatial-audio}  presents the learning procedure, and includes the AVC and AVSA learning methods, the visual crop selection procedure and spatial audio processing, while Section \ref{sec:binaural} presents the stereo audio setup. Section \ref{sec:experimental} presents the experimental setup and results, including the dataset, the AVC and AVSA tasks results, the human action recognition video-only tasks, and the audio-only Ambisonics and binaural acoustic scene classification.
Finally Section \ref{sec:concl} presents conclusions and future work.

\section{Related work}
\label{sec:related}

Self-supervised learning from audio-visual data has provided a number of powerful models for a variety of downstream tasks. In those methods, firstly, proxy learning tasks are framed as correspondence or coincidence between the audio and visual signals. The methods are then commonly tested on downstream tasks such as  single-modality classification problems, e.g., acoustic scene or sound classification, image classification, human action recognition from video, etc. Below we review the approaches most related to ours.

One of the most well-known such methods is Look, Listen and Learn, or L\textsuperscript{3}-Net, \cite{DBLP:journals/corr/ArandjelovicZ17}, which explores what can be learnt from a large number of unlabelled videos using the video and audio information without any supervision. To achieve this, authors proposed an AVC learning task by training a video subnetwork and an audio subnetwork simultaneously to predict if a frame of a video corresponds or not to an audio segment. Positive pairs come from the same video where 1-second audio clip overlap the video frame. Negative pairs are obtained by randomly choosing two different videos, picking one random frame from one video and a random 1-second audio segment from the other. The authors found that the method could solve the AVC task, and also learn good visual and audio features. The effectiveness of audio features was tested on two sound classifications tasks \cite{piczak2015esc}, \cite{stowell2015detection} and set new state-of-the-art performance at the time, while visual features evaluated on the ImageNet challenge \cite{russakovsky2015imagenet} also performed on par with the state-of-the-art methods. 

More recent work by Cramer et al. \cite{8682475} studied the effect of audio embeddings obtained from various audio input representations to L\textsuperscript{3}-Net, showing that mel-frequency log-magnitude spectrograms used as the input to the audio subnetwork outperform the linear-frequency log-magnitude spectrograms. In addition, the authors also found that it is not necessary to match the content of audio for training embeddings to the audio in the downstream tasks, which further indicates that L\textsuperscript{3}-Net is able to learn more general and robust features. 
The approach was used for audio-visual scene classification as a baseline system in the DCASE 2021 Challenge \cite{Wang2021}. Audio features and visual features were extracted from the pretrained audio and video subnetworks of L\textsuperscript{3}-Net, respectively, after which the audio and video embeddings were concatenated into a single feature representation. Classification using the joint feature representation was found to outperform significantly the single-modal embeddings, but more advanced image-based approaches outperformed the given baseline system in the challenge \cite{Wang2021}.

While methods in \cite{DBLP:journals/corr/ArandjelovicZ17,8682475} have focused on learning to match the video and audio streams, in \cite{yang2020telling}, the authors proposed to learn representations by learning a spatial correspondence task between video and audio which aligns the spatial information in the audio towards the given positions in the video frame. Unlike the works in \cite{DBLP:journals/corr/ArandjelovicZ17,8682475} in which the system is trained to tell if a video frame corresponds or not to an audio segment, the work proposed by Yang et al. \cite{yang2020telling} is designed to distinguish if a video frame spatially aligns with an audio frame. Positive pairs are created by using a video with its original audio left-right channels, while negative pairs are generated by flipping the video's left-right audio channels. The effectiveness of the learnt representation was evaluated on  downstream tasks such as sound localization, audio spatialization, and audio-visual source separation. The authors showed that by learning such spatial correspondence, the system achieves quantitative gains over those that do not leverage spatial audio cues.

Similar to \cite{yang2020telling}, Morgado et al. \cite{NEURIPS2020_328e5d4c} also emphasize the importance of spatial cues which often occur in audio and video streams. To learn from the spatial information, authors trained a network to distinguish if crops in a 360\textdegree \ video frame are spatially aligned with the corresponding ambisonic audio. It should be noted that previous works have focused on training a two-stream network (video stream and audio stream) with a final linear layer which does the binary classification task to tell if a video frame corresponds or not to an audio frame \cite{DBLP:journals/corr/ArandjelovicZ17}, or if a video frame spatially aligns or not with an audio frame \cite{yang2020telling}, where binary cross entropy loss is used to regularize the network. However, in \cite{NEURIPS2020_328e5d4c}, the authors adopted contrastive losses \cite{NEURIPS2020_d89a66c7,oord2018representation,chen2020simple} which are prevalent in recent work in the vision community to perform audio-visual spatial alignment. Given a batch of video-audio pairs, features from a positive pair are pulled together while, for a negative pair, are pushed away.

To solve this audio-visual spatial alignment task, the authors also adopted a curriculum learning strategy \cite{bengio2009curriculum} by dividing the task into two stages. Firstly, the network is trained to identify AVC at the instance level where a positive pair is obtained from a crop sampled from a random angle in a video frame and its corresponding rotated ambisonic audio of the same video, and a negative pair is created from a crop sampled from a random angle in a video frame and the rotated ambisonic audio from a different video from the same batch. In the AVSA stage, the contrastive learning is done at both instance-level and crop-level, where four crops from a video frame and their corresponding rotated ambisonic audio from the same video form positive pairs, and four crops from a video frame with their misaligned Ambisonics from the same video and other videos form negative pairs, which also means that a crop in a video frame is distinguished against the other three crops in the same video frame and the other crops in different videos. The effectiveness of the features was evaluated on both in-domain downstream tasks such as AVC and AVSA, and out-of-domain downstream tasks such as video segmentation and action recognition. 

Our work builds upon the above-mentioned method, and proposes  selecting object-oriented crops instead of random viewing angles, adopting explicit spatial audio features instead of just multichannel mel-spectrograms, and investigating the effect of different combinations between object-oriented crops and spatial features. We also investigate the effectiveness of the learnt audio embeddings in audio-only downstream tasks, using stereo and mono audio, something that the authors of \cite{NEURIPS2020_328e5d4c} were not concerned with.

\section{Learning from spatial audio features}
\label{sec:spatial-audio}

\begin{figure*}
    \centering
    \includegraphics[scale = 0.8]{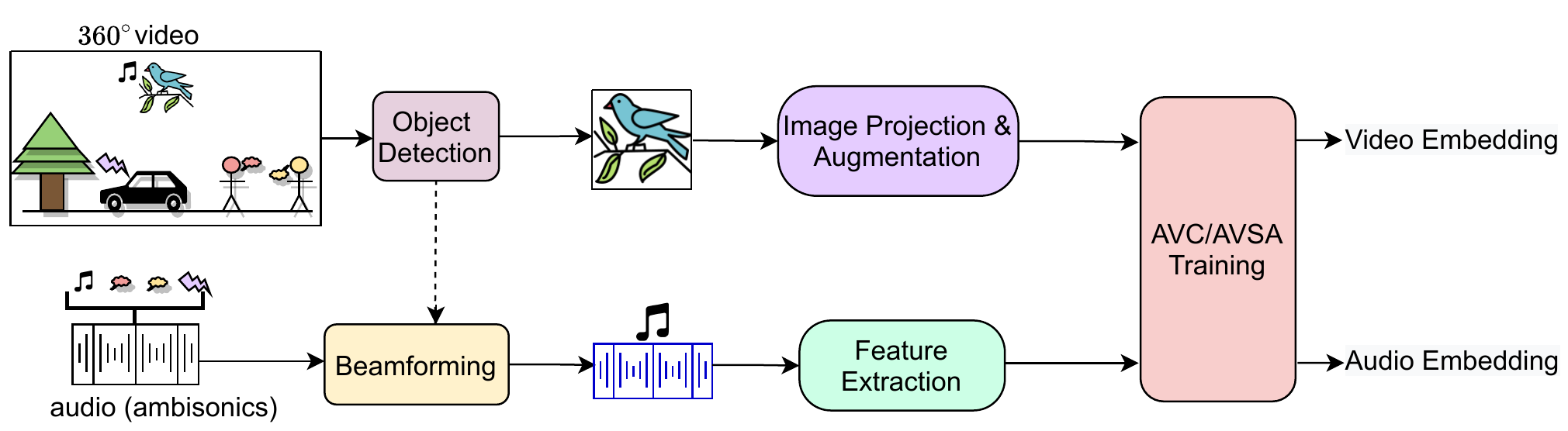}
    \caption{Simplified block diagram of the proposed system. Object-oriented crops are selected from the video using an object detection method, and the ambisonic audio is rotated towards the center of the crop to form positive pairs.}
     \label{fig:concept}
     \vspace{-10pt}
\end{figure*}

A simplified block diagram of the proposed method is illustrated in Fig. \ref{fig:concept}. A video clip and corresponding audio in the training dataset are first processed before being input into the learning system, which is either AVC or AVSA, depending on the used learning procedure. 
The AVC learning process aims to learn feature representations based solely on the correspondence (the temporal overlap) of the audio and the video clip. On the other hand, the AVSA aims to learn feature representations by using multiple crops and rotated audio signals of the same clip, and their spatial correspondence (temporal overlap and correct spatial orientation). Once trained, the system will be able to produce as its output video and audio embeddings for the video frames and audio representation presented as its input.

\begin{figure*}
    \centering
    \includegraphics[scale = 0.9]{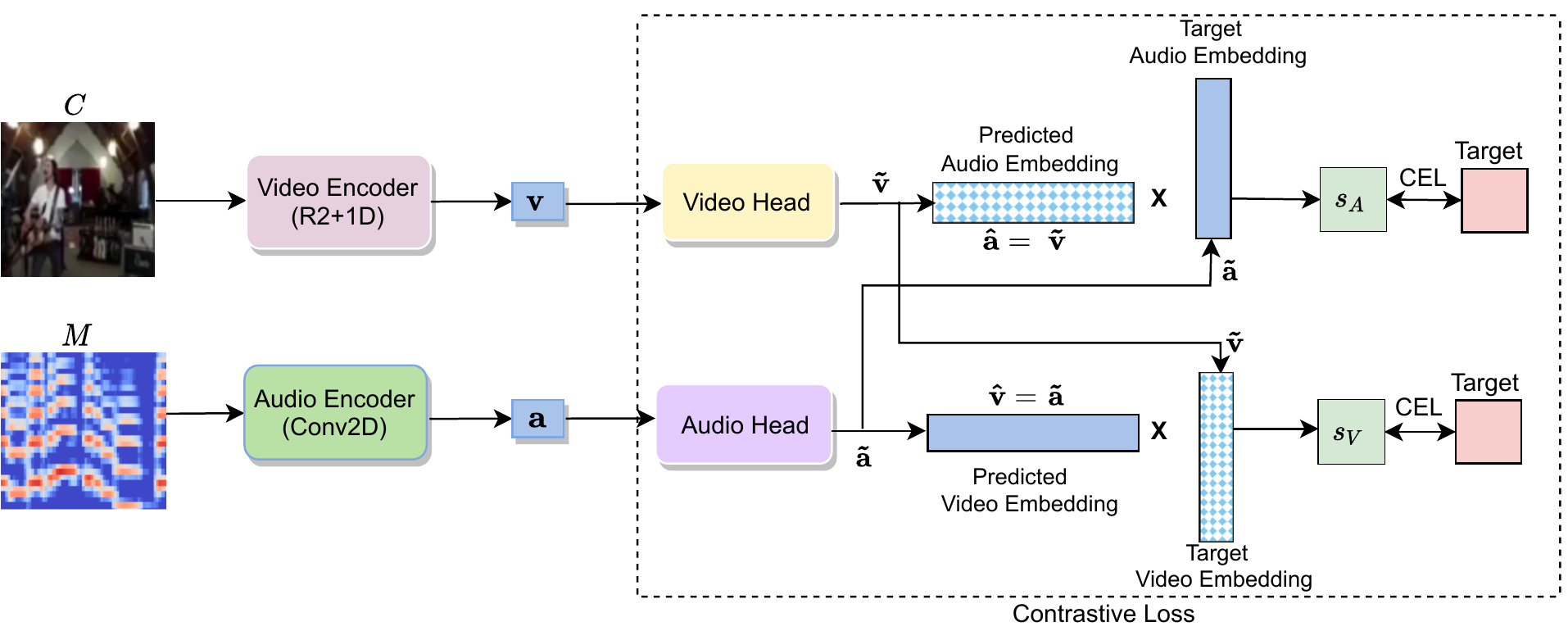}
    \caption{Block diagram of the AVC learning. For each audio-video pair, one crop and its corresponding correct or rotated ambisonic audio are used.
    }
    \label{fig:AVC-block-diag}
    \vspace{-10pt}
\end{figure*}

\subsection{Data preprocessing}

To select audio-video pairs, 0.5 s of video clip with 16 fps resulting in eight frames, and 1 s of audio with 24 kHz sampling rate which overlaps with the video are chosen.
For each frame, the objects in the image are detected using YOLO \cite{redmon2016-yolo} applied on the equirectangular image (360\textdegree \ image), returning objects with their bounding boxes and center points. From the detected objects, one crop is randomly chosen, and the ambisonic audio signals are rotated or beamformed towards the selected object crop direction, as will be explained in Section \ref{sec:spatial-audio}.

The object crops are further processed by applying a gnomonic projection in order to transform the equirectangular frame into normal field-of-view crops given the center point.\footnote{https://github.com/NitishMutha/equirectangular-toolbox} Crops are further resized to 112 × 112 resolution. Standard image augmentation techniques are applied to increase the variability of the training data: RandomHorizontalFlip which flips the images horizontally from left to right with a probability 50\%, and ColorJitter which randomly changes the brightness, contrast, saturation and hue of the images.

The audio is processed through a feature extraction step, which extracts either log-mel or spatial features, as will be explained in
Section III. For the versions with multiple channels, log-mel spectrograms are extracted from each channel and are stacked along the channel axis. Spectrograms are calculated using a  window size of 21 ms with a hop size of 10 ms, resulting in 100 frames for a 1s segment of audio, further mapped onto the mel scale using 128 mel bands. 

The processed crops from the video signal and feature representation of the audio signal, obtained as explained above, form the pairs which are used as input to the video and audio encoders, respectively, and which drive the AVC or AVSA learning process.

 \subsection{Audio-visual correspondence learning}

The block diagram of the AVC learning procedure is presented in Fig. \ref{fig:AVC-block-diag} for processing a single clip selected from the training dataset. For AVC, one crop $C$ corresponding to an object detected in the image is selected, and the audio features $M$ of the corresponding rotated audio are used per clip. 
We follow the same architecture used in \cite{NEURIPS2020_328e5d4c} to have a fair comparison. The system consists of two encoders: one video encoder which adopts an 18-layer R2+1D model (a 3D convolution, implemented as 2D followed by 1D, and used within a ResNet architecture) as  proposed in \cite{8578773} and one audio encoder which is a 9-layer 2D convolutional neural network (CNN). 
The object crops, and the corresponding spectral audio features
are input into the video and audio encoders, respectively. These encoders produce the video embedding $\mathbf{v}   \in \mathcal{R}^{512}$ and audio embedding $\mathbf{a}   \in \mathcal{R}^{512}$. 

Video embeddings are projected to target video embeddings $\mathbf{\tilde{v}} \in \mathcal{R}^{128}$ by a video head, and the audio embeddings are projected to target audio embeddings $\mathbf{\tilde{a}} \in \mathcal{R}^{128}$ by an audio head. The target video embeddings are also used as predicted  audio embeddings and vice versa, so that $ \hat{\mathbf{a}} = \mathbf{\tilde{v}}$ and $\hat{\mathbf{v}} = \mathbf{\tilde{a}}$. 
This setting helps audio and video embeddings from the positive pairs to  come closer and negative pairs to be apart. Both video and audio heads consist of one linear layer of 128 neurons.

For a batch of $N$ audio-video pairs, the cosine similarity is calculated between the predicted audio  $\hat{\mathbf{a}}_i$ outputted from the video head and the target audio ${\mathbf{\tilde{a}}}_j$ outputted from the audio head, where $i$ and $j$ are the clip index, $i,j=1,...,N$. At the same time, the cosine similarity is calculated between the predicted video $\hat{\mathbf{v}}_i$ outputted from the audio head, and the target video $\mathbf{\tilde{v}}_j$ outputted from the video head:
 \begin{equation}
 \textrm{sim}(\hat{\mathbf{a}}_i, \mathbf{\tilde{a}}_j)   = \frac{\hat{\mathbf{a}}_i \cdot {\mathbf{\tilde{a}}_j}} {\|\hat{\mathbf{a}}_i\|\|\mathbf{\tilde{a}}_j\|} \hspace{6pt}\textrm{and} \hspace{6pt}
  \textrm{sim}(\hat{\mathbf{v}}_i, \mathbf{\tilde{v}}_j)   = \frac{\hat{\mathbf{v}}_i \cdot {\mathbf{\tilde{v}}_j}} {\|\hat{\mathbf{v}}_i\|\|\mathbf{\tilde{v}}_j\|}
 \end{equation}

Finally, cross-entropy loss (CEL) is applied between each cosine similarity matrix and the labels, with the target in the CEL calculation being 1 for audio and crops that originate from the same clip (i.e. $i=j$), and 0 for for audio and crops that originate from different clips (i.e. $i \neq j$). 
In practice, this means that,    for positive audio-video pairs, the video subnetwork is trained to generate embeddings as similar as possible to the audio embeddings generated by the audio subnetwork and vice versa.  
The two subnetworks are trained jointly. 

\subsection{Audio-visual spatial alignment learning}

\begin{figure*}
    \centering
    
    \includegraphics[width=\textwidth]{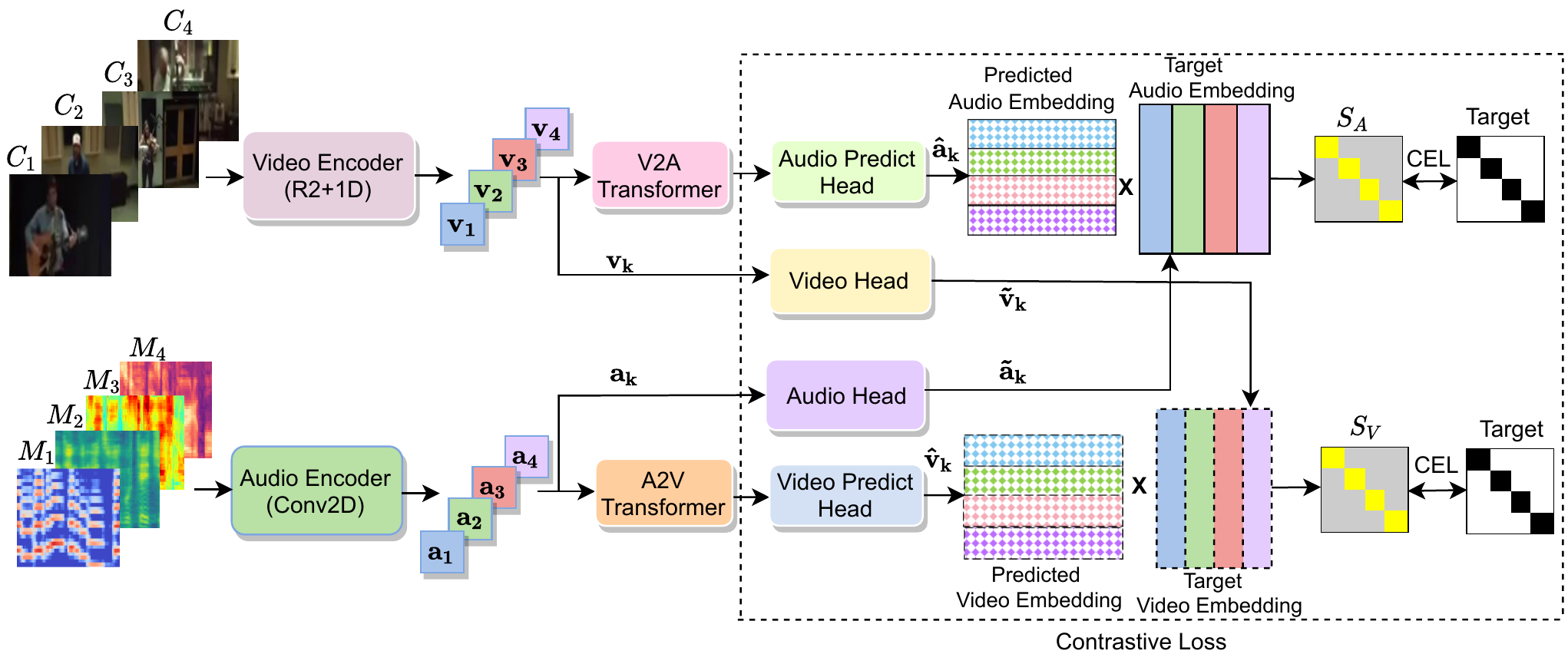}
    \caption{Block diagram of the AVSA learning. Four crops and their corresponding rotated ambisonic audio are used for each audio-video pair.}
   \label{fig:avsa}
\end{figure*}

\begin{figure}
    \centering
    \includegraphics[width=\linewidth]{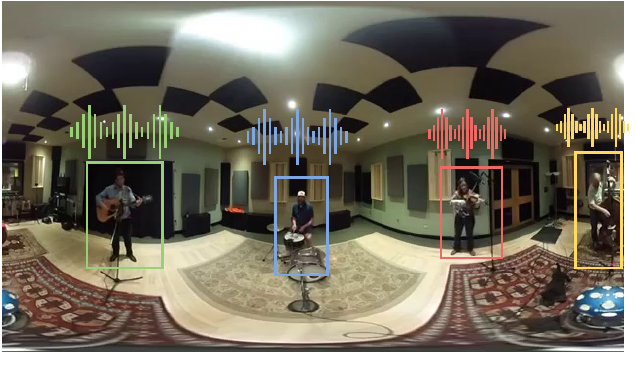}
    \caption{The AVSA learning procedure uses four crops of the same clip and corresponding audio, presented as green, blue, red and yellow pairs.}
    \label{fig:audio_video pair}
    \vspace{-10pt}
\end{figure}

Audio-visual spatial alignment is a difficult task, and implemented by dividing it into two stages. The first stage is AVC, as explained above, in which the system is trained to identify  correspondence at the instance level. Second, in the AVSA stage, the system is trained to identify correspondence both at instance-level and crop-level. 

The AVSA learning procedure follows the same general procedure as the AVC learning, but instead of taking one crop for each frame, four crops and their corresponding audio are taken for each clip.
The block diagram of AVSA is presented in Fig. \ref{fig:avsa} for processing a single clip selected from the training dataset. Four crops $C_{k}$ ($k = 1,2,3,4$) are selected randomly or based on the objects detected in the image, and the corresponding rotated audio features $M_{k}$ are computed, centered at each crop, as illustrated in  Fig. \ref{fig:audio_video pair}. For each crop and rotated audio, embeddings are produced by the video and audio encoder with the same configuration as in AVC stage, respectively. 

The audio embeddings $\mathbf{a}_{1},...,\mathbf{a}_{4}  \in \mathcal{R}^{512}$ output from the audio encoder, are further projected by an audio head into lower dimension, producing the target audio embeddings $\mathbf{\tilde{a}}_{1},...,\mathbf{\tilde{a}}_{4}   \in \mathcal{R}^{128}$. At the same time, the audio embeddings are translated into video features of the same size using a translation network. This translation network between two modalities, the A2V transformer, is similar to the transformer of \cite{NIPS2017_3f5ee243}, and is necessary for ensuring the accurate translation of features for the projection step that follows. This is because audio clips contain the signal from all the listening angles while only those video crops falling into the given angle could be seen. After the A2V transformer, the representation is further projected by the video predict head into lower dimension, producing the predicted video embeddings $\hat{\mathbf{v}}_{1},...,\hat{\mathbf{v}}_{4}  \in \mathcal{R}^{128}$. 

The same steps are performed on the video branch side: the video embeddings $\mathbf{v}_{1},...,\mathbf{v}_{4}  \in \mathcal{R}^{512}$ output from the video encoder are projected by a video head into the target video embeddings $\mathbf{\tilde{v}}_{1},...,\mathbf{\tilde{v}}_{4}  \in \mathcal{R}^{128}$, and are also translated to audio features through V2A transformer, then projected by the audio predict head into the predicted audio embeddings $\hat{\mathbf{a}}_{1},...,\hat{\mathbf{a}}_{4} \in \mathcal{R}^{128}$.
All the four projection heads consist of a linear layer of 128 neurons. For details of the transformers, we refer our readers to \cite{NEURIPS2020_328e5d4c}.

For a batch of $N$ clips, the cosine similarity is calculated between the predicted audio embeddings $\hat{\mathbf{a}}_{ik}$ and the
target audio embeddings $\mathbf{\tilde{a}}_{jl}$, similarly between the predicted video
embeddings $\hat{\mathbf{v}}_{ij}$ and the target video embeddings $\mathbf{\tilde{v}}_{ij}$, where $i$ and $j$ are the clip index $i,j =1,...,N$ and $k$ and $l$ are the crop index for each clip $k,l =1,...,4$.
  \begin{equation}
 \textrm{sim}(\hat{\mathbf{a}}_{ik}, \mathbf{\tilde{a}}_{jl})   = \frac{\hat{\mathbf{a}}_{ik} \cdot {\mathbf{\tilde{a}}_{jl}}} {\|\hat{\mathbf{a}}_{ik}\|\|\mathbf{\tilde{a}}_{jl}\|} \textrm{ and } 
  \textrm{sim}(\hat{\mathbf{v}}_{ik}, \mathbf{\tilde{v}}_{jl})   = \frac{\hat{\mathbf{v}}_{ik} \cdot {\mathbf{\tilde{v}}_{jl}}} {\|\hat{\mathbf{v}}_{ik}\|\|\mathbf{\tilde{v}}_{jl}\|}
 \end{equation}
Cross-entropy loss (CEL) is applied between the cosine similarity and the labels. In this case there are multiple pairs per clip, and the network must learn to differentiate within-clip and between-clip pairs. The target in the CEL calculation is 1 for audio and crops that originate from the same clip and are spatially aligned ($i=j$ and $k=l$) , and 0 for for audio and crops from the same clip but misaligned ($i=j$ and $k \neq l$), and also for audio and crops from different clips ($i \neq j$). 

It should be noted that, in AVC learning, the object crop and its corresponding rotated audio in one audio-video pair are contrasted with the ones from a different audio-video pair in a batch.  However, in AVSA learning, each crop with its corresponding audio in one audio-video pair is contrasted with different crops in the same pair,  and with crops from other audio-video pairs. This process allows learning of the spatial alignment within the same clip.

\subsection{Spatial selection of video crops with YOLO}

\begin{figure}
    \centering
    \includegraphics[scale = 0.7]{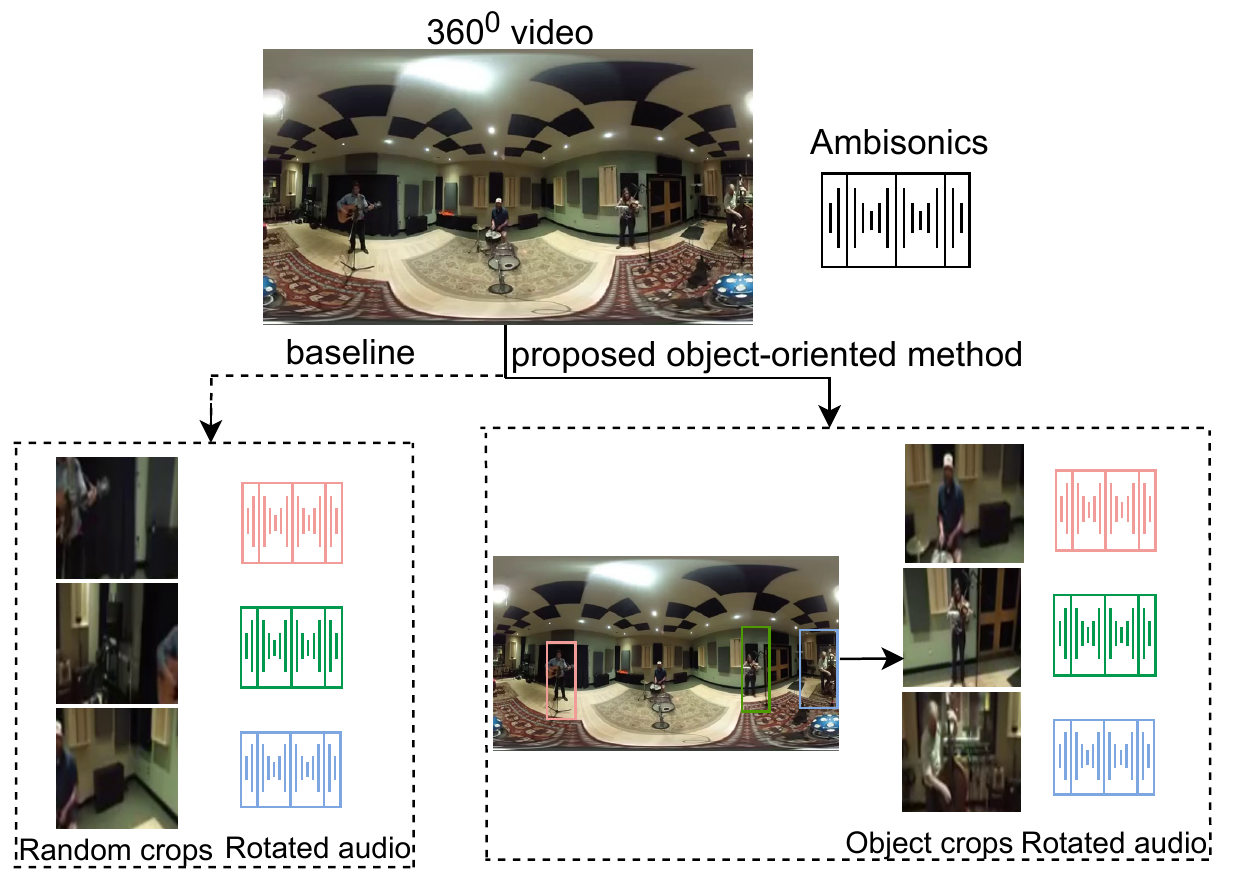}
    \caption{Comparison of the procedure for choosing image crops between baseline and proposed method.     }
    \label{fig:block-compare}
    \end{figure}

Using random crops with random field of view can increase the variability of the dataset in terms of information presented to the network in the learning process. However, there can be many crops with very little information content, such as crops with only dark background, small objects, or tiny parts of a person's dress. In addition, visual objects are most likely to correspond to sound sources. For example, a car passing by on the road is often heard along with the engine noise; people in the meeting room tend to create speech. Alignment that focuses on learning from the direct sound of the source and the corresponding visual object are likely to produce a stronger correspondence.
To do this, we propose to select crops which are object-oriented, using YOLO to detect the objects \cite{redmon2016-yolo}. 

YOLO detector is  applied on the equirectangular frames (360\textdegree \ image) of the video directly, returning objects with their bounding boxes and center points.  In AVC learning, one crop is randomly chosen from the detected objects. If no objects are found by YOLO, one random crop is chosen the same way as in \cite{NEURIPS2020_328e5d4c}. In the AVSA training, four crops are chosen from those detected crops, having center points falling into azimuth angle $[180,90]$ (left-back), $[90,0]$ (left-front), $[0,-90]$ (right-front), $[-90,-180]$ (right-back), respectively. If no crop is detected by YOLO within the target angle area, a random crop is taken for that area. This procedure ensures that the four crops are apart from each other, to avoid redundant information. 

The comparison of the crop selection method for the two approaches is presented in Fig. \ref{fig:block-compare}.

\subsection{Spatial sound focusing with beamforming}
The effect of different spatial audio formats, including mono channel audio, was investigated in \cite{NEURIPS2020_328e5d4c} in the ablation studies. The authors tested three cases: a \emph{mono} case, a \emph{stereo} case, and the \emph{ambi} case using the full ambisonic signal set. Both the \emph{mono} and \emph{stereo} cases were generated from the ambisonic signals as described further in the text. The reference to a \emph{mono} signal can be confusing, since in typical audio recording terminology it would commonly refer to a single-channel recording without directional selectivity, i.e. an omnidirectional recording, which corresponds to the first channel of Ambisonics. However, by \emph{mono} the authors of \cite{NEURIPS2020_328e5d4c} actually refer to a beamforming operation using the ambisonic signals, steering a beam towards each of the video crops. The process can be formulated as follows. The first-order ambisonic (FOA) format has a directional response to a wave incident from azimuth and elevation angles $(\theta,\phi)$
\begin{equation}
 \mathbf{u}(\theta,\phi) = \left[ \begin{array}{c}
    1\nonumber\\
    \sin\theta\cos\phi\nonumber\\
    \sin\phi\nonumber\\
    \cos\theta\cos\phi
    \end{array} \right]
    \label{eq:ambi_format}
\end{equation}
where the three lower elements correspond to the Cartesian components of a unit vector pointing to the direction-of-arrival (DOA). The format can also be interpreted as an omnidirectional channel and three directional channels with dipole directivities oriented along the three principal axes.
For a source signal $s(n)$ then, the ambisonic signals $\mathbf{x}(n)$ are
\begin{equation}
    \mathbf{x}(n) = \left[ \begin{array}{c}
        w(n) \\
        y(n) \\
        z(n) \\
        x(n) 
    \end{array} \right] = \mathbf{u}(\theta,\phi)s(n).\label{eq:ambi_src_enc} 
\end{equation}
It is evident that the FOA format encodes directly the DOA information of the source in the resulting signals. Capturing of a general scene of multiple source signals, diffuse ambience and reverberation can be formulated through the concept of a plane-wave amplitude density $a(\theta,\phi, n)$, describing a continuous distribution of incident waves. The FOA signals for such a general scene are then
\begin{equation}
    \mathbf{x}(n) = \int_{-\pi}^\pi \int_{-\pi/2}^{\pi/2} 
    \mathbf{u}(\theta,\phi)
    a(\theta,\phi, n) \cos\phi \mathrm{d}\phi \mathrm{d}\theta
    \label{eq:ambi_scene_enc}
\end{equation}

Beamforming with ambisonic signals is typically done using a weighted version of the encoding basis as beamforming weights \cite{rafaely2021fundamentals}. In the simplest case, the beamformed signal $y(n)$ for a beamforming direction $(\theta_0,\phi_0)$ is
\begin{equation}
    y(n) = \mathbf{u}^\mathrm{T}(\theta_0,\phi_0)\mathbf{x}(n)\label{eq:ambi_beam}.
\end{equation} 
Another common operation in Ambisonics is the capability to rotate the sound field. In the case of FOA only, and contrary to higher-order Ambisonics, the rotation can be simply performed with a standard rotation matrix. Details on constructing such matrices can be found in \cite{politis2016jsAmbisonics}. Following e.g., the yaw-pitch-roll convention corresponding to angles $(\alpha,\beta,\gamma)$ such a rotation $\mathbf{Q}$ is applied to the ambisonic signals as
\begin{equation}
    \mathbf{x}_\mathrm{rot}(n) = \mathbf{Q}(\alpha,\beta,\gamma)\mathbf{x}(n).\label{eq:ambi_rot}
\end{equation}

\begin{figure}[t]
    \centering
    \includegraphics[width = \columnwidth]{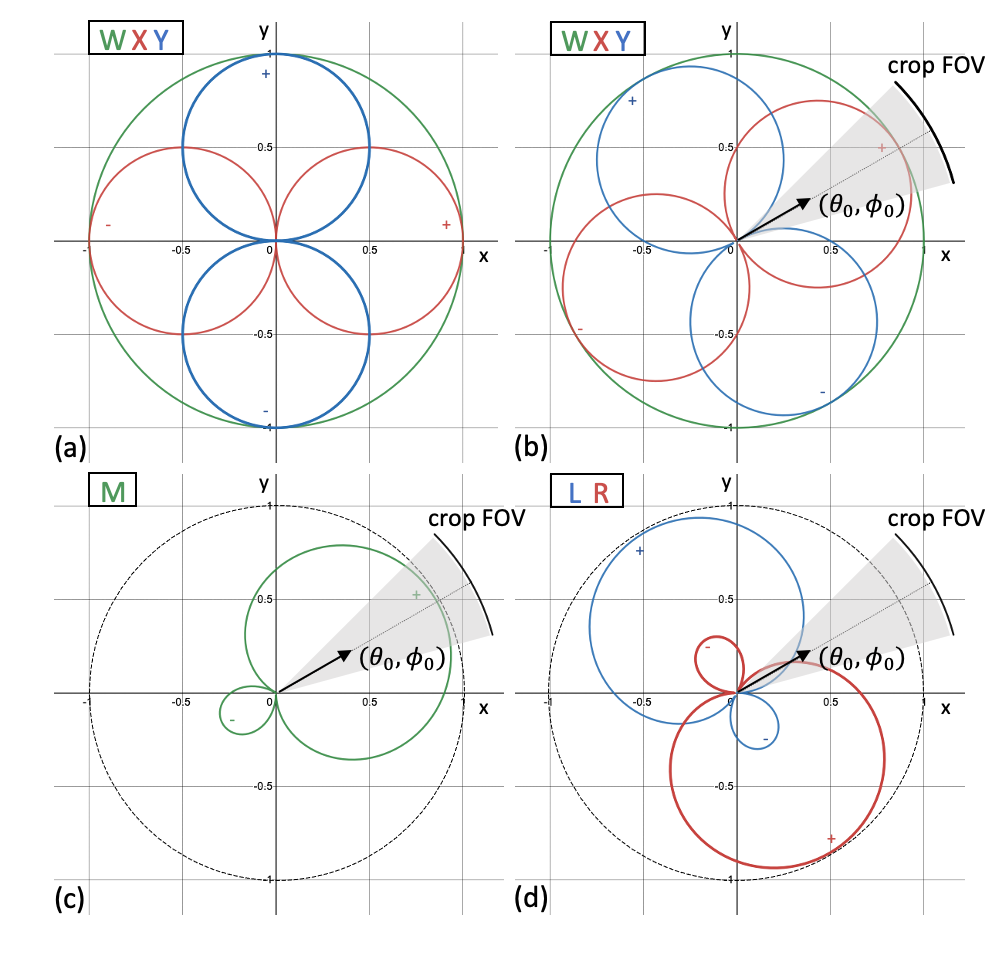}
    \caption{Spatial characteristics of the spatial formats tested in AVC and AVSA. Horizontal only view omitting the \emph{Z} vertical channel of FOA. The signs indicate polarity of the directional responses for the respective lobes. (a) original FOA format (omitting Z channel), (b) rotated FOA towards the crop center, (c) mono format from beamforming towards crop center, (d) stereo format from two left-right beamformers around crop center.}
    \label{fig:FOA}
    \vspace{-10pt}
\end{figure}

In the case of \cite{NEURIPS2020_328e5d4c}, the \emph{mono} case was generated by applying Eq.~\ref{eq:ambi_beam} with the beamforming direction oriented at each crop center $(\theta_0,\phi_0)$. Extraction of the \emph{stereo} is detailed in the following section. The full \emph{ambi} case is generated by rotating the whole ambisonic recording in order to align its frontal axis with the crop center, using a $\mathbf{Q}(\theta_0,-\phi_0,0)$ rotation matrix. The directional responses of the channels of each case for a certain crop direction are depicted in Fig.~\ref{fig:FOA}.

Tested on four downstream tasks, experiments showed that direct beamforming as input has no advantages over \emph{stereo} or \emph{ambi}, so the authors of \cite{NEURIPS2020_328e5d4c} concluded that a single channel per crop may not be sufficient to represent spatial relations between sounds in the scene. However, since the crops are randomly taken from a video frame, the beams may be focusing in regions where there are no sound sources present matching the visual information, therefore the audio signal may be dominated by ambience generated by all visible and non-visible sources in the scene. On the other hand, even if the beamformed signal captures a strong source contribution from the beamformed direction, the random visual crop may not be spatially aligned with it, isolating a patch from some non-informative background instead. To minimize such audiovisual mismatches, we implement the beamforming combined with the object-oriented crops obtained from YOLO detector. The motivation behind this idea comes from the fact that human beings tend to look for the objects in a scene which are producing the sounds. 
 
\subsection{Spatial audio features}

In the work of \cite{NEURIPS2020_328e5d4c} spatial information from audio is passed to the network in the form of 4-channel mel-band energies $\mathbf{x}^\mathrm{mel}_{tf} = [w^\mathrm{mel}_{tf},\; y^\mathrm{mel}_{tf},\; z^\mathrm{mel}_{tf},\; x^\mathrm{mel}_{tf}]^\mathrm{T}$, where $(t,f)$ indicate time and mel-band indices respectively. Mel-band energies are suitable to partially express inter-channel level differences, such as those naturally occurring between ambisonic signals captured with Eq.~\ref{eq:ambi_src_enc}, or between the left and right stereo channels in a recording, captured with directional microphones or produced using stereo panning. However, in the case of Ambisonics some directional information is lost with mel-band energies since signal polarity (sign) differences are discarded, resulting in non-unique values of inter-channel level differences for more than one direction. 

A more appropriate spatial feature for ambisonic signals is the active intensity vector (AIV), an acoustical quantity indicating the mean flow of sound energy, which has a long history in sound source localization \cite{pavlidi20153d} and spatial audio coding \cite{pulkki2017first}. Recently, it has been used in deep learning-based source localization \cite{perotin2019crnn} and it proved to be the most popular spatial audio feature in the DCASE sound even localization and detection challenge \cite{politis2020overview} with ambisonic input. In this work we are using a normalized version of the AIV as in \cite{perotin2019crnn}, which bounds the magnitude of the vector between [0,1], similar to the \emph{diffuseness} feature \cite{pulkki2017first}, with unity length in the presence of a single source, and less than unity in the presence of multiple sources or noise and ambience. The normalized AIV is given by
\begin{equation}
    \mathbf{i}_{tf} = \frac{2 \Re\left( w_{tf}^* \left[ \begin{array}{c}
    x_{tf}\\
    y_{tf}\\
    z_{tf}
    \end{array}\right]\right)}{|w_{tf}|^2+|x_{tf}|^2+|y_{tf}|^2+|z_{tf}|^2},\label{eq:aiv}
\end{equation}
with $\Re(\cdot)$ being the real part of a complex number.
Note that the AIV is computed with the original STFT complex spectrograms. To end up with similar time-frequency dimensions as the mel-band energies $\mathbf{x}^\mathrm{mel}$, the AIV features are aggregated across mel-bands similar to \cite{Cao2019}.
Finally, training including those features is done by stacking them as additional channels along with the mel-band energies 
\begin{equation}
    \mathbf{x}^\mathrm{sf}_{tf} = \left[ \begin{array}{c}
        \mathbf{x}^\mathrm{mel}_{tf}\\
        \mathbf{i}_{tf}
        \end{array}
    \right].\label{eq:spat_feat_ambi}
\end{equation}
The cases where such features are used for training are indicated as \emph{FOA-IV} in the results.

\section{Learning from stereo audio}
\label{sec:binaural}

Apart from beamforming in the \emph{mono} case, 2-channel stereo input in the \emph{stereo} case is additionally tested. While training for spatial alignment using \emph{stereo} is found better than using \emph{mono}, it is still suboptimal against the full \emph{ambi} case. However, it is of interest to investigate further the performance of training for stereo signals in downstream tasks since there are many more audio-visual datasets with stereo audio than with FOA audio, and it is significantly easier to collect new ones using commodity cameras providing stereo signals, than FOA. Morgado et. al. in \cite{NEURIPS2020_328e5d4c} avoid this issue by focusing on the performance of the learned audio-visual features in video-only downstream tasks. However, if the method is to be applied to audiovisual tasks, or audio-only tasks, extension to mono or stereo input is important. 

Stereo signals corresponding to coincident stereo recording \cite{zotter2019Ambisonics} can be extracted in a straightforward way from FOA signals, by generating two beamformers emulating two coincident directional microphones pointing left and right. Note that ambisonic-to-binaural decoding can be also conducted in the same manner, with frequency-dependent beamformers approximating generic or individualized head-related transfer functions \cite{politis2016jsAmbisonics}. In this work we use the same process as in \cite{NEURIPS2020_328e5d4c}, where two broadband hyper-cardioid beamformers are steered towards $90^\circ$ to the left and right respectively, falling somewhere between stereophonic recording and binaural decoding. Since the \emph{stereo} case is extracted after alignment of the sound scene with the crop center, obtaining the stereo signals $\mathbf{y}_\mathrm{st}(n) = [l(n),\; r(n)]^\mathrm{T}$ combines the beamforming of Eq.~\ref{eq:ambi_beam} and rotation of Eq.~\ref{eq:ambi_rot}
\begin{equation}
    \mathbf{y}_\mathrm{st}(n) = [\mathbf{u}(90^\circ,0), \; \mathbf{u}(-90^\circ,0)]^\mathrm{T}\mathbf{Q}(\theta_0,-\phi_0,0)\mathbf{x}(n).
\end{equation}

Similar to the FOA features, the stereo signals are converted into log mel-band energy spectrograms $\mathbf{y}^\mathrm{mel}_{tf} = [l_{tf}^\mathrm{mel},\, r_{tf}^\mathrm{mel}]^\mathrm{T}$, where $l_{tf},r_{tf}$ are the left and right channels accordingly. In addition to the log mel-band energies, we extract certain spatial features suitable for stereo and binaural signals \cite{wang2018combining} such as the \emph{inter-channel level differences} 
\begin{equation}
    d_{tf} = \log\left(\frac{|l_{tf}|^2}{|r_{tf}|^2}\right) = 2\log|l_{tf}| - 2\log|r_{tf}|
\end{equation}
and \emph{inter-channel phase differences}
\begin{equation}
    \pmb{\phi}_{tf} = \left[ 
    \begin{array}{c}
        \cos(\angle l_{tf}-\angle r_{tf}) \\
        \sin(\angle l_{tf}-\angle r_{tf}) 
    \end{array}\right]
\end{equation}
where $\angle$ refers to extracting the argument of a complex number. Similar to the spatial features for ambisonic signals \emph{FOA-IV}, the stereo spatial features are aggregated in mel-bands to match the spectrogram dimensions of $\mathbf{y}^\mathrm{mel}_{tf}$ and stacked as additional channels
\begin{equation}
    \mathbf{y}^\mathrm{sf}_{tf} = \left[ \begin{array}{c}
        \mathbf{y}^\mathrm{mel}_{tf}\\
        d_{tf}^\mathrm{mel}\\
        \pmb{\phi}_{tf}^\mathrm{mel}
        \end{array}
    \right].\label{eq:spat_feat_stereo}
\end{equation}
The  cases  where  such  features  are  used  for  training  are indicated as \emph{ICF} in the results.

\section{Experimental results}
\label{sec:experimental}

We evaluate the proposed extensions on the in-domain tasks of AVC and AVSA that are the same as used for training the system. We also apply our method to the purely visual-based downstream task of human action recognition, in order to allow a comparison to the method in \cite{NEURIPS2020_328e5d4c}, namely the one where training is based on randomly selected image crops and the feature representation for audio consists of the log-mel spectrum, to which we will refer to as the baseline system. We then concentrate on audio downstream tasks. 

\subsection{Dataset and data format verification}

Training is based on the YouTube-360 dataset that was collected from YouTube and contains 360\textdegree \ video with spatial audio. The dataset contains a total of 5506 videos, of which 4506 videos are meant for training and 1000 for testing. Automatic curation of content was performed to avoid silent regions. The videos were segmented into clips of 10 s, and only clips with a volume level over a certain threshold were used, resulting in 88733 clips (246 hours of content). The dataset is unlabeled. A more detailed description of the dataset collection procedure and curation is presented in \cite{NEURIPS2020_328e5d4c}. 

Since the proposed approaches depend on the spatial encoding of Ambisonics, and since such encoding is integrated into the audio signals themselves and not, e.g., additional metadata, it is crucial to know exactly which channel ordering and channel normalization convention is used, in order to perform beamforming and rotation operations without errors. In this work we used the widely accepted ambisonic convention of \emph{ACN} channel ordering, which corresponds to (WYZX) order for FOA, and \emph{SN3D} channel normalization scheme, together known as \emph{ambiX} format \cite{politis2016jsAmbisonics}. However since Ambisonics are not defined yet in audio file container formats, confusion with other surround formats and channel swapping by audio file writers and loaders is a common occurrence. For example, downloading the YT-360 dataset from YouTube in the AAC format results in a 6-channel file with two additional channels that may or may not be used for an additional orientation-independent stereo track\footnote{\url{https://github.com/google/spatial-media/blob/master/docs/spatial-audio-rfc.md}}. To verify the correct channel ordering with the current downloading and audio loading sequence before processing the recordings for feature extraction, we uploaded a test file with known channel order to YouTube, downloaded it, reloaded it using \emph{pyAV}\footnote{https://github.com/PyAV-Org/PyAV}, and identified the correct final channel mapping to WYZX. We then applied this mapping during loading to the whole YT-360 dataset.

Beyond standardizing the channel order in the dataset, there is still a high chance that a number of files do not conform to the ambisonic format specifications. The reasons for this may be many: a) the user did not upload an ambisonic file but a stereo file with extra empty channels, or some other surround format, b) the user uploaded an ambisonic file with the wrong channel order, c) the recording device or the encoding software provided a badly encoded ambisonic file to the user. All such cases will result in erroneous alignment operations in this work, which may affect the result. Due to the diversity of media sources in YouTube, and the newness of the ambisonic format in consumer media recording and production, it is expected that a high number of files in the dataset may be invalid. To investigate that aspect further, a simple test was devised to indicate how close a recording is to an ideal ambisonic recording. The test is formulated as
\begin{equation}
    \abs*{\frac{E_\mathrm{xyz}}{E_\mathrm{w}}-1} \leq \tau
\label{eq:test} 
\end{equation}
where $E_\mathrm{w} = \sum_n w(n)^2$ is the energy of the omnidirectional channel, $E_\mathrm{xyz} = \sum_n (x(n)^2+y(n)^2+z(n)^2)$ is the total energy of the three dipole channels, and $\tau<1$ is a threshold value. The test is based on the principle that for any combination of uncorrelated sources in the sound scene, and incoherent diffuse ambience and reverberation, due to the orthogonality of the encoding in Eq.~\ref{eq:test} the omnidirectional signal energy should be equal to the total energy captured by the three dipole signals. Since ambisonic microphones and recording devices suffer from spatial aliasing at higher frequencies which compromises the encoding, we apply a low-pass filter to all signals below 4 kHz before applying the test, below which we assume that most devices can deliver accurate ambisonic encoding. Using a threshold of $\tau=0.1$ only approximately 28\% of files pass the test in both the training and the test set\footnote{The list of clips that pass the test can be found at \newline https://github.com/shanwangshan/YT-360-Ambisonics}. Note that the test is strict in the sense that there can be valid combinations of correlated sources coming from different directions in the scene, which can violate the test, but such cases are less common in reality.

\begin{table}
    \small
    \centering
    \begin{tabular}{l|c|c}
    \midrule
    Training data  & AVC task  & AVSA task\\
      
     \midrule
    100\%, all data         &  90.97 &   71.75\\
    28\%, all Ambisonics     &  80.43 &  59.84 \\
    28\% all not Ambisonics  & 87.25  &  60.41 \\
       \bottomrule
    \end{tabular}
    \vspace{1mm}
    \caption{Accuracies (\%) of AVC and AVSA downstream tasks using different subsets of the training dataset. The test dataset is complete for all cases.}
    \label{tab:ambi_or_not}
    \vspace{-16pt}
\end{table}

Surprised by the result described above, we evaluate the outcome of the AVC and AVSA training procedure in three cases: using the complete training subset, using only the 28\% of files from the training set that pass the test in Eq.~ \ref{eq:test}, and using randomly selected 28\% of files from the training set that do not pass the test. No selection of the clips is done for the test set, therefore it contains both strictly ambisonic data and not ambisonic. For this experiment we use the baseline system setup in which the crops are selected at random and the audio is represented using log-mel features. The results for the AVC and AVSA training and downstream tasks are presented in Table \ref{tab:ambi_or_not}.

The AVC task seems to benefit from having a variety of erroneously rotated data, as there is enough inconsistency and variety for the system to learn to be invariant to the orientation and just learn the correspondence. This behavior  is similar to the contrastive learning in \cite{fonseca2021unsupervised} in which positive pairs were different views of the data created through augmentation, and the system learned the correspondence in a manner that is invariant to the transformation applied by the augmentation method to the data. The strictly ambisonic data is more homogenous in spatial content, therefore the smaller amount of data results in lower performance. On the AVSA task, the smaller data amounts perform similar whether ambisonic or not, but with lower performance than the entire, larger, dataset. 
Since in practice deviations from ideal ambisonic encoding are expected in datasets captured from diverse sources, also reflected in the test set of YT-360, it seems that even the AVSA task can benefit from such examples included during training.
In the following experiments, we use the entire dataset when training.

\subsection{Audio-visual correspondence and spatial alignment}

We first evaluate the proposed extensions on the in-domain tasks of AVC and AVSA that are the same as used for training the system. The results for these tasks are presented in Table \ref{tab:avc-avsa}, starting with the baseline system as proposed in \cite{NEURIPS2020_328e5d4c}, and with all the additional elements proposed in this study. Unless mentioned, the feature representation used for the audio signal is the log-mel spectrogram. While the AVC and AVSA tasks are the same as what the method is trained on, we observe a significant improvement brought by the FOA-IV features. In particular, the spatial alignment benefits of the use of spatial features, as predicted, with an absolute increase in performance of 10\%.
YOLO improves on the AVC results, but not on AVSA, while beamforming on its own does not bring any advantage to any task. The combination of beamforming and YOLO also does not seem to bring any advantage for the in-domain tasks when used with the mel features, while the FOA-IV features with YOLO have a significant effect in improving performance. 

When it comes to the stereo audio, the system performs on the AVC task significantly worse than with Ambisonics, while in the AVSA there is no significant drop in performance. Stereo audio with YOLO appears to worsen performance in both AVC and AVSA tasks, whether with mel features or inter-channel features. 

\begin{table}[]
    \small
    \centering
    \begin{tabular}{l|c|c}
    \midrule
    & AVC task & AVSA task \\
        \midrule
    Baseline \cite{NEURIPS2020_328e5d4c} & 90.97  & 71.75 \\
    \midrule
    FOA + YOLO & \textbf{92.73} & 70.83\\
    FOA + FOA-IV & \textbf{91.56} & \textbf{81.06}\\
    FOA + FOA-IV + YOLO & 91.18 & 80.97\\
    \midrule
    Stereo & 84.67 & 71.69\\
    Stereo + YOLO & 82.22 & 69.62\\
    Stereo + YOLO + ICF & 82.37 & 69.15\\
    \midrule
     Beamform & 80.60 & 62.68\\
    Beamform +YOLO & 80.91 & 63.29\\   
    \bottomrule
    \end{tabular}
    \vspace{1mm}
    \caption{Accuracies (\%) of different methods tested on AVC and AVSA downstream tasks. Values in bold are outside of the 95\% confidence interval w.r.t the baseline performance.}
    \label{tab:avc-avsa}
    \vspace{-16pt}
\end{table}

Based on these results, we can clearly state that the use of spatial information in the feature representation of the audio signal results in better learnt representations for the in-domain tasks. Changing the random crops to YOLO on the video signal also improves slightly, but significantly, the AVC task performance. On the other hand, training with stereo audio does not benefit from YOLO or spatial features for the AVC and AVSA tasks.

\subsection{Human action recognition}

\begin{table*}[]
    \small
    \centering
    \begin{tabular}{l|c|c|c|c}
    \midrule
    & \multicolumn{2}{c|}{without fine tuning} & \multicolumn{2}{c}{ with fine tuning}\\
     Dataset &  UCF   & HMDB  &  UCF  & HMDB \\  
            
     \midrule
    
    Baseline \cite{NEURIPS2020_328e5d4c} & 36.3  & 23.1  & 65.9  & 32.0 \\
    \midrule
    FOA + YOLO & 35.7 & 23.0 & 65.0  & 32.9 \\
    FOA + FOA-IV & 36.9  & 22.7 & \textbf{67.6}  & 33.3 \\
    FOA + FOA-IV + YOLO &  37.5  &24.0   & 66.2  &32.4 \\
    \midrule
    Stereo & 37.1  & 23.4  & 65.6 & 32.0  \\
    Stereo + YOLO & 40.1  & \textbf{26.2} & 66.9  & \textbf{34.2}  \\
    Stereo + YOLO + ICF & 40.1 &25.1  & 65.7  & 30.7 \\
    \midrule
    Beamform & 37.1  & 22.1 & 65.9  & 31.8 \\
    Beamform +YOLO & \textbf{40.4}  & 24.2  & 66.9  & \textbf{34.2} \\    
    \bottomrule
    \end{tabular}
    \vspace{1mm}
    \caption{Accuracies (\%) of action recognition downstream task without and with fine tuning. }
    \label{tab:action-rec}
    \vspace{-10pt}
\end{table*}

The downstream task of human action recognition is based on two different datasets, the UCF dataset \cite{journals/corr/abs-1212-0402} that contains 101 classes, and the HMDB dataset \cite{6126543} that contains 51 classes. The classification is based on the video embeddings only, as these datasets do not have any audio content. 

A simple way to test the quality of the learned features is to use them as embeddings and append one simple linear layer of 101 neurons (UCF dataset), or 51 neurons (HMDB) for performing the classification. While the parameters of the classification layer are trained for the task, the weights of the pretrained network remain fixed. 
Another approach to downstream  tasks is to allow updating the parameters of the entire network, namely the linear classification layer and all the layers of the pretrained structure, in a process called fine tuning, which will compensate some of the domain mismatch between the proxy and downstream tasks and data. 

We compare the proposed method with the baseline system both with and without fine tuning. The results are presented in Table \ref{tab:action-rec}, for the system trained using the AVSA proxy task. The results are presented at clip level, i.e. one clip per video, following the procedure in \cite{NEURIPS2020_328e5d4c}.

Without fine tuning, our proposed additions to the baseline system outperform the baseline in most cases on the UCF dataset, with the highest performance being obtained when using both beamforming and YOLO in training by 4\% improvement. Curiously, even the training with stereo audio and YOLO outperforms the baseline, even though the downstream task makes no use of the audio embeddings. Similarly, for the HMDB dataset, which is much smaller in size, the embeddings based on the stereo audio with YOLO are outperforming significantly the baseline by 3\%. 
When using fine tuning, the advantage brought by the different extensions is diminished, even though the performance obtained is, in some cases, higher. On UCF, the method using FOA-IV performs the best, with 1.7\% advantages over baseline; on HMDB, the method using beamforming with YOLO and stereo with YOLO achieve the highest with 2.2\% advantages over baseline. In both cases their performance lies in the 95\% confidence interval of the baseline.

The above results show that the use of spatial information in the audio signal brings important benefits to the learning process, which has an effect not only on the in-domain tasks that use audio (as shown in the previous section), but also results in superior video embeddings, as demonstrated by their effect on the human action recognition task. The advantage of the spatial information in audio is preserved also for the stereo format: the system using stereo audio, YOLO and ICF performs significantly better than the baseline when used without fine-tuning.

\subsection{Acoustic scene classification with ambisonic audio}

For testing the quality of the audio embeddings resulting from the learning process, we use them on a new downstream task, acoustic scene classification using ambisonic audio. For this task, we use the Eigenscape dataset \cite{app7111204}, consisting of 640 minutes of audio recordings belonging to eight classes: Beach, Busy Street, Park, Pedestrian Zone, Quiet Street, Shopping Centre, Train Station, and Woodland. The data was recorded using the mh Acoustics Eigenmike with a windshield to prevent wind noise, and further converted to Ambisonics. Recordings were cut into 30-s segments, and a cross-validation setup is provided with the data for comparison of methods. The spatial classification system using Directional Audio Coding (DirAC)-based features \cite{pulkki2017first} provided with the data in \cite{app7111204} had a 69\% performance, later outperformed by a CNN approach, with 82\% performance \cite{green2019acoustic}.

Our results are presented in Table \ref{tab:eigenscape}. The system does not use any fine tuning, training only a linear classification layer for the task. 
The classification performance when using the AVC training is 82.5\%, while for the AVSA training it is 89.4\%, showing the superiority of the learned representations when spatial correspondence is considered. While the AVC proxy training attains a similar performance to the previous SOTA on this data, the AVSA proxy training outperforms it significantly.
The use of FOA-IV instead of the log-mel representation does not bring any improvement for the AVC-trained system, and is even detrimental to classification performance on the AVSA-trained system. 

\begin{table}
    \small
    \centering
    \begin{tabular}{l|c|c}
    \midrule
     
      & AVC training & AVSA training\\
      
      \midrule
      Baseline \cite{NEURIPS2020_328e5d4c} & 82.5 &\textbf{89.4}\\
          FOA-IV & 82.5 & 74.1\\
   
    \bottomrule
    \end{tabular}
    \vspace{1mm}
    \caption{Accuracies (\%) of acoustic scene classification on EigenScape dataset, no fine tuning} 
    \label{tab:eigenscape}
    \vspace{-16pt}
\end{table}

\subsection{Acoustic scene classification with binaural audio}

We test a two-channel downstream audio classification tasks, for which we train the system using stereo audio with the AVSA pretext task, as this has shown the highest performance in the previous experiments. The quality of the learned features is first tested without fine-tuning, using the pretrained network as a feature generator. We also use it with fine-tuning in order to investigate the highest achievable performance on this task. 

While there is a mismatch between the stereo format that we used for training the system and the binaural format of the audio data we use for testing, we expect the method to work reasonably well. The main difference is the effect of true head-related transfer functions (HRTFs) in the binaural format, which include strong inter-aural time differences at low-mid frequencies and frequency-varying level differences at mid-high frequencies, while the stereo version there contains only level differences that are mostly frequency-independent for the range of valid ambisonic encoding.

\begin{table*}
    \small
    \centering
    \begin{tabular}{l|c|c|c|c|c|c}
    \toprule
     
     & \multicolumn{3}{c|}{without fine tuning} &  \multicolumn{3}{c}{with fine tuning}\\
      & Audio only & Video only  & Early A-V fusion & Audio only  & Video only & Early A-V fusion\\
      
      \midrule
      Open L3 system \cite{Wang_ICASSP2021} & \textbf{75.8} & \textbf{68.4} & \textbf{82.2} & - & - & -\\
      \midrule
      Beamforming & 62.9 & 55.9 & 74.7 & 72.3 & 76.9 & 80.5\\
      Beamforming + YOLO &  62.6 & 54.7 & 72.7 &  69.2 & 73.3 & 79.5\\
      \midrule
      Stereo & 63.4 & 54.5 & 72.9 & 69.2 & 73.7 & 81.1 \\
    Stereo + YOLO &  64.0 & 54.1 & 72.6 &  68.2 & \textbf{78.1} & \textbf{81.3}\\
    Stereo + YOLO + ICF & 65.0 & 53.2 & 73.9 & 68.3 & 72.2 & 76.8 \\
    \bottomrule
    \end{tabular}
    \vspace{1mm}
    \caption{Accuracies (\%) of Audio-Visual Scene Classification on TAU Audio-Visual Urban Scenes 2021 development dataset.}
    \label{tab:asc-binaural}
    \vspace{-10pt}
\end{table*}

The experiment is based on the TAU Audio-Visual Urban Scenes 2021 dataset \cite{Wang_ICASSP2021}, which consists of ten classes, 34 hours of audio and synchronized video, recorded in ten different large European cities. Recording of audio was performed with binaural in-ear microphones, to represent the audio information as it would be received by a human listener, while the video was recorded using a GoPro camera mounted on the strap of the backpack, offering a frontal view of the scene. The person recording the data was requested not to move while recording, therefore the position of the recording setup was fixed for each clip. 

We investigate the classification accuracy for the system using audio-only, video-only and early fusion of audio and video modalities. The results are presented in Table \ref{tab:asc-binaural}, and compared with the baseline system provided with the dataset \cite{Wang_ICASSP2021}. The baseline uses OpenL3 \cite{cramer2019look} for producing audio and video embeddings, and uses them directly as features. Its accuracy is 75.8\% and 68.4\% for the audio-only and video-only classification, respectively, while early fusion, which concatenates the audio and video embeddings in a single feature vector, obtains an accuracy of 82.2\%.

In comparison, the system proposed in this work has a significantly lower performance, as can be seen in Table \ref{tab:asc-binaural}. The lower performance is seen in both audio-only and video-only classification, irrespective of the use of YOLO and ICF. However, using fine-tuning boosts significantly the effect of YOLO on the video side for the stereo format,  resulting in around 5\% improvement compared with the stereo format without YOLO, and a 10\% increase compared with the video-only performance of the OpenL3 system. Early fusion in our system has similar performance with the OpenL3-based on (within 95\% CI), except the version using ICF; ICF seem to be detrimental to learning robust video embeddings, which result in lower performance of the video-only and the fusion system. 

Based on this particular test case, it seems that for the task of audio-visual scene classification the embeddings produced with OpenL3 are better than the ones obtained with the proposed system. We note that there are important differences  between the two approaches: different datasets, different  network architectures, different optimization mechanisms and loss functions in learning, and not least, different proxy tasks. We do however believe that the AVSA learning may have advantages for tasks that target spatial information, as shown by the high performance in the AVSA task itself. Acoustic scenes contain a prevalent background, and the AVSA following foreground sounds and objects may not have sufficient additional information in order to affect the performance.

\section{Conclusions and future work}
\label{sec:concl}

In this work, we proposed a self-supervised learning method for learning audio representations based on spatial alignment between audio and video information. To create a strong correspondence between the audio and video content, we proposed a new method for sampling crops by detecting the objects in the video frame using YOLO. Additionally, to use the spatial audio information from Ambisonics to its full extent, we proposed use of acoustic intensity vector as feature representation for the audio input. Our results show that by using YOLO only, the AVC performance improves 1.7\%, which is statistically significant given the size of the test data (15116 clips). The use of FOA-IV instead of conventional log-mel spectrogram boosts the AVSA performance by 10\%, showing that indeed FOA-IV model the spatial information in the ambisonic audio better. The use of FOA-IV has a strong effect on the overall AVSA learning process.

We also investigated the different combinations of audio input features with YOLO, namely YOLO wih beamforming and with FOA-IV. 
We found that by using the combination of YOLO and beamforming improves the action recognition task. The advantage of this method is particularly strong in action recognition tasks without fine tuning, 
obtaining a 4\% improvement on the UCF dataset and 1.2\% increase on the HMDB dataset. This reinforces our assumption that by selecting an object and focusing on the sound that specific object makes can potentially result in better feature representations.

The effectiveness of the audio embeddings learned through AVSA was also tested on acoustic scene classification using ambisonic audio from the EigenScape dataset. We found a 7\% advantage for AVSA training compared with AVC training, which further validates the authors' \cite{NEURIPS2020_328e5d4c} conclusion, but from an audio embedding point of view.

Since audio with ambisonic format is still not very prevalent, we investigated learning of audio embeddings using mono and stereo format, and tested on TAU Audio-Visual Urban Scenes 2021 dataset for scene classfication. We found that both learnt audio and video embeddings are underperforming compared to the embeddings produced using OpenL3 as used in \cite{Wang_ICASSP2021}. However, with the help of fine tuning, the performance of audio model, video model and early A-V fusion model get boosted compared with no fine tuning. Especially, a significant boost of 24\% was seen for the video model only with stereo+YOLO method after fine tuning, which surpasses the Open L3-based method by 10\%. Further investigation using different tasks is needed to assess the performance on audio data that has more directional information than acoustic scenes, for example sound event classification, localization, or tagging, in which foreground sounds are dominant.

For future work, we consider that a more sophisticated detection method which works on equirectangular images is worth investigating for sampling of the crops as done in \cite{8546070}. This is because YOLO is trained on normal images, and might not give the optimal detection when directly applied on equirectangular images. Another interesting direction would be selecting areas of interest based on directions of prominent sound activity, using more advanced techniques for acoustic imaging from FOA, e.g.  direct-path time-frequency selection \cite{nadiri2014localization}.
Additionally, the current learning scheme is to produce one embedding vector for each second of audio instead of each analysis frame, which limits  applicability of the learnt audio embeddings to classification, while many audio-related downstream tasks require a higher temporal resolution, e.g., sound event detection or sound event localization. In future, this learning scheme could be modified to cater for a wider variety of audio-related downstream tasks.

\bibliographystyle{IEEEtran}
\bibliography{references}

\begin{IEEEbiography}[{\includegraphics[width=1in,height=1.20in,clip,keepaspectratio]{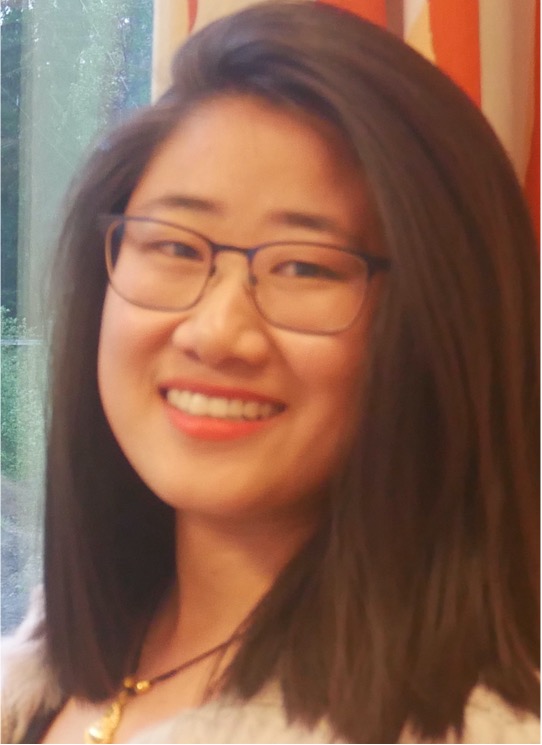}}] {Shanshan Wang} is a doctoral student at Tampere University since 2020 in the field of Computing and Electrical Engineering. She has received her MSc. degree in Information Technology in 2019 at Tampere University, and BSc. degree in Information and Computing Science in 2017 at China University of Petroleum (East China), China. Her research focuses on computational scene analysis using audio-visual data and she is interested in audio signal processing, self-supervised learning and multi-model analysis.
\end{IEEEbiography}

\begin{IEEEbiography}[{\includegraphics[width=1in,height=1.20in,clip,keepaspectratio]{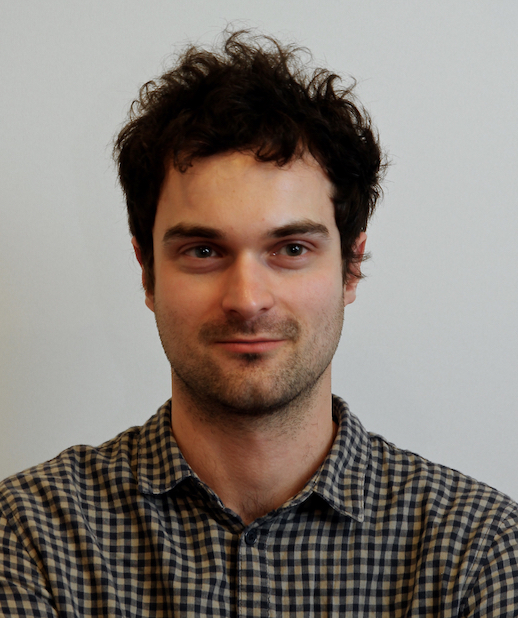}}]{Archontis Politis}
is a post-doctoral researcher at Tampere University, Finland. He obtained his M.Eng. degree in civil engineering from Aristotle University, Thessaloniki, Greece, and his M.Sc. degree in Sound \& Vibration studies from the Institute of Sound and Vibration Research (ISVR), Southampton University, UK, in 2006 and 2008, respectively. 
In 2016 he obtained a Doctor of Science degree on parametric spatial sound recording and reproduction from Aalto University, Finland. 
His research interests include spatial audio technologies, virtual acoustics, array signal processing, and acoustic scene analysis.
\end{IEEEbiography}

\begin{IEEEbiography}[{\includegraphics[width=1in,height=1.20in,clip,keepaspectratio]{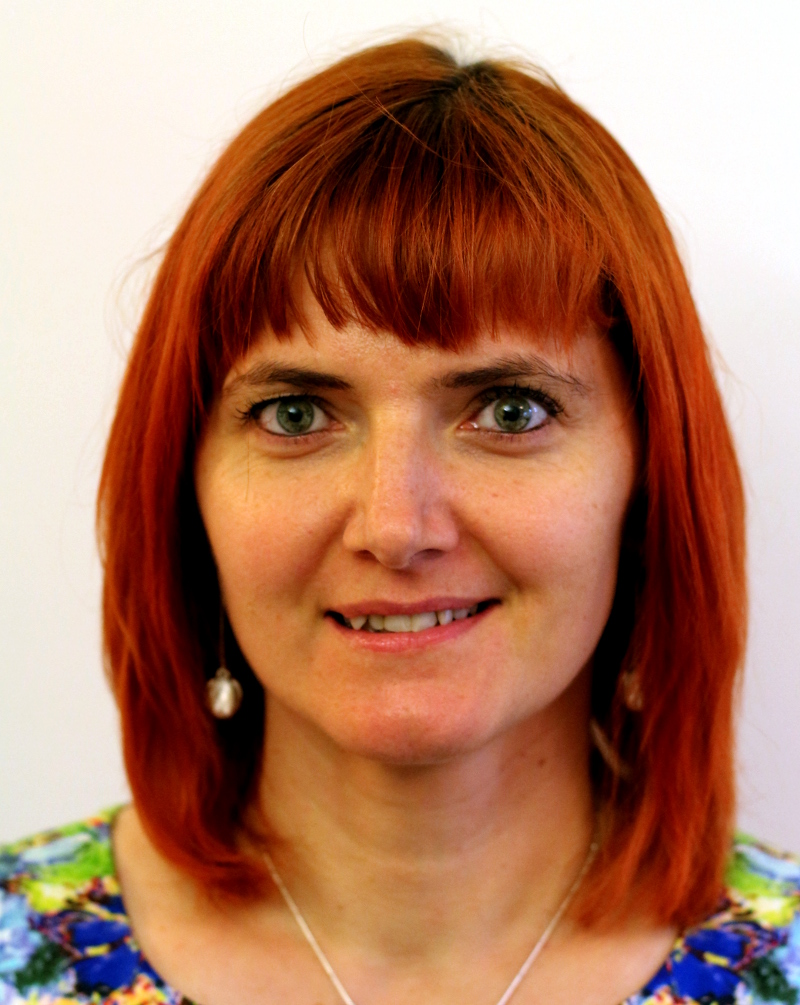}}]
{Annamaria Mesaros} is Assistant Professor at Tampere University. She received her PhD in Signal Processing at Tampere University of Technology in 2012. Her research focuses on sound event detection in real-world multisource environments,
and includes over 35 scientific publications and many open datasets. She is coordinator of the Detection and Classification of Acoustic Scenes and Events (DCASE) Challenge, and member of the Audio and Acoustic Signal Processing Technical Committee of IEEE Signal Processing Society.
\end{IEEEbiography}

\begin{IEEEbiography}[{\includegraphics[width=1in,height=1.2in,clip,keepaspectratio]{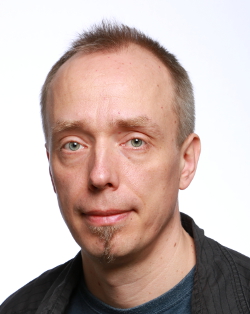}}]{Tuomas Virtanen} 
is Professor at Tampere University. He received his doctoral degree from Tampere University of Technology in 2006. He is known for his pioneering work on single-channel sound source separation using nonnegative matrix factorization and computational analysis of sounds in everyday environments. He has authored over 200 scientific publications on these topics, and has received the IEEE Signal Processing Society 2012 best paper award. He is an IEEE Fellow, and recipient of the ERC 2014 Starting Grant "Computational Analysis of Everyday Soundscapes".
\end{IEEEbiography}

\vfill

\end{document}